\documentclass[twocolumn,showpacs,amsmath,amssymb]{revtex4-1}
\usepackage{graphicx}
\usepackage{dcolumn}
\usepackage{bm}
\usepackage{amssymb}
\usepackage{color}
\begin{document}
\newcommand{\hesan}{$^3$He }
\newcommand{\heyon}{$^4$He }
\newcommand{\kake}{$\times$}
\newcommand{\aone}{A$_1$ }
\newcommand{\atwo}{A$_2$ }
\newcommand{\tc}{$T_C$ }
\newcommand{\tcone}{$T_{C1}$ }
\newcommand{\tconeEnd}{$T_{C1}$}
\newcommand{\tctwo}{$T_{C2}$ }
\newcommand{\tctwoEnd}{$T_{C2}$}
\newcommand{\tconstant}{$\tau$ }
\newcommand{\tone}{T$_1$ }
\newcommand{\toneinv}{$T_1^{-1}$ }

\title{Spin Fluid Dynamics Observed by Magnetic Fountain Effect and Mechano-Spin Effect in the Ferromagnetic Superfluid $^3$He \aone Phase}
\author{Y. Aoki}\altaffiliation[Present address: ]{Department of Material Science and Engineering, Tokyo Institute of Technology, Nagatsuka, Midori Yokohama 226-0026 Japan}
\author{A. Yamaguchi}\altaffiliation[Present address: ]{Graduate School of Material Science, University of Hyogo, Kamigori Hyogo  678-1297 Japan.}
\author{K. Suzuki}
\author{H. Ishimoto}
\affiliation{Institute for Solid State Physics, Kashiwa, Chiba 123-456 Japan}
\author{H. Kojima}\affiliation{Serin Physics
Laboratory, Rutgers University, Piscataway NJ 08854}

\begin{abstract}
Systematic observations of the magnetically generated fountain pressure in the superfluid \hesan \aone have been carried out in a newly built apparatus designed to reduce the effect of thermal gradients.  In the same apparatus, mechanical pumping and filtering of polarized nuclear spins were realized by the pneumatic pumping action of an electrostatically actuated membrane.  In both experiments, the measured induced pressure was observed to decay at all temperatures where the \aone phase appeared in magnetic fields up to 13 T and liquid pressures between 1 and 29 bar.  The inferred spin relaxation rate tended to increase as the low temperature phase boundary with the \atwo phase (\tctwo) was approached.  The increase in spin relaxation rate near \tctwo can be explained by the presence of a minority spin condensate in the \aone phase as predicted by Monien and Tewordt and by the application of the Leggett-Takagi theory of spin relaxation in superfluid $^3$He.
\end{abstract}


\maketitle

\section{introduction}
The superfluid phases of liquid \hesan appear below a pressure dependent transition temperature $T_C$.  In zero applied magnetic field two phases exist and are known as the A (having spin pairings in opposite directions but equal energy gaps: $\Delta_{\uparrow\uparrow} = \Delta_{\downarrow\downarrow}$) and the B (having all spin pairings with energy gaps, $\Delta_{\uparrow\uparrow}$, $\Delta_{\downarrow\downarrow}$ and $\Delta_{\uparrow\downarrow}$) phases.  The A phase occurs only at relatively high pressures and temperatures and shares a triple point in a pressure-temperature phase diagram with the B and the normal phases near 21 bar.  Under applied magnetic field, the phase diagram sheds the triple point and acquires a new \aone phase between two transition temperatures $T_{C1}$ and $T_{C2}$ (where $T_{C2}<T_C<T_{C1})$) at all pressures \cite{Wheatley75,Leggett75}.
The \aone phase has been regarded as a ``ferromagnetic" superfluid phase whose condensate only involves totally spin polarized pairs with the energy gap $\Delta_{\uparrow\uparrow}\neq 0$ but $\Delta_{\downarrow\downarrow}\equiv 0$ \cite{Ambegaokar73}.  Unique magneto-hydrodynamics \cite{Liu79} of the \aone phase leads to such effects as the spin-entropy waves \cite{Corruccini80,Bastea04}, the magnetic fountain effect (MFE) \cite{Ruel85,Yamaguchi06}, spin-current induced electric fields \cite{Hu82} and the excitation of spin and mass supercurrents via the Aharonov-Casher effect \cite{Balatsky93}.  The latter two have not yet been observed experimentally to our knowledge.  Under an applied magnetic field and below $T_{C2}$ $\Delta_{\downarrow\downarrow}$ begins to grow in the phase known as $A_2$. The magnetohydrodynamics effects listed above are absent in the \atwo phase.

Although the spin fluid dynamics of the \aone phase have been studied over many years, there still remain outstanding questions \cite{Kojima08}.  The most important among these questions is the origin of the unexpected spin relaxation observed in MFE experiments.  Understanding this spin relaxation would yield important clues in designing a spin pumping device for boosting the spin polarization to much greater level than feasible by available static magnetic fields \cite{Yamaguchi09b}.  In this report, we describe our recent experiments on both magnetically and mechanically driven spin superflows in the superfluid \hesan \aone phase.  In these experiments, the underlying principle for observing the spin superfluid dynamics is the mechanical detection of the spin density variation.  Measurements were made on the \aone phase under pressures between 1 and 29 bar and static magnetic fields up to 13 T.  Improvements to the apparatus have eliminated the thermal gradients that marred an earlier experiment and precluded us from acquiring an accurate measure of the temperature dependence of the spin relaxation.  From an analysis of the measured decay of the pressure change induced by applying a magnetic field gradient or by spin pumping, the spin relaxation time (T$_1$) is extracted.  The extracted \tone decreases monotonically as the temperature is lowered and tends to vanish at $T_{C2}$.  This behavior is \emph{unexpected} for a totally ferromagnetic superfluid \aone phase but is consistent with the \aone phase containing a small amount of minority spin pair condensate \cite{Yamaguchi07}. The presence of minority condensate is in agreement also with the theoretical predictions of Monien and Tewordt \cite{Monien85}.

This report is organized as follows.  In section II, the magnetic fountain and spin pumping effects are described in terms of simple two fluid model equations applied to our experimental apparatus.  Spin relaxation is incorporated into this model phenomenologically, so as to account for the characteristic timescale
over which induced spin pressures are observed to decay.  In section III, the details of the apparatus are described.  In section IV, results and analyses are presented and the paper concludes with a summary in Section V.

\section{two fluid hydrodynamics of magnetic fountain and mechano-spin effects}
Consider a small detector chamber enclosed except for an opening to the narrow channels of an attached superleak.  The superleak channels connect the detector interior to a large reservoir volume.  The detector chamber, superleak channels and the reservoir volume are all filled with liquid He$^3$.  The flow impedance of the superleak is such that the normal (N) fluid flow is severely restricted by both the large shear viscosity of normal liquid \hesan at low temperatures and the large flow impedance of the long narrow opening in the superleak structure.  One wall of the detector chamber is a stretched flexible membrane. The differential pressure between the interior of the detector chamber and the reservoir produces a membrane deflection which can be measured.  In the quasistatic limit (where the superfluid acceleration is negligible), the superfluid maintains equality of chemical potential across the ends of the superleak \cite{Liu79}:
\begin{equation}\label{fountainEffect}
      \frac{\delta P}{\rho} = \frac{\gamma\hbar}{2m}\left[ \delta H - \frac{\gamma}{\chi}\delta S\right],
\end{equation}
where $\delta P \equiv P_R - P_D$, $\delta S \equiv S_R - S_D$ and $\delta H \equiv H_R - H_D$ are the differential pressure, spin density and external magnetic field, respectively, along the superleak channels between the reservoir($R$) and detector($D$) ends, $m$ is the mass of \hesan atom, $\rho$ is the mass density, $\gamma$ is the magnitude of the gyromagnetic ratio, and $\chi$ is the magnetic susceptibility.

If a magnetic field gradient is applied across the superleak in an ideal arrangement where the membrane deflection is negligible ($\delta S = 0$), a pressure gradient is developed according to Eq.~(\ref{fountainEffect}). This is the MFE \cite{Liu79}.  It is also possible to apply a pressure gradient across the superleak in the absence of magnetic field gradient.  In this case Eq.~(\ref{fountainEffect}) implies that a spin density gradient should result.  We call this a mechano-spin effect(MSE) \cite{Dombre82}.

Since the volume of the reservoir is much larger than that of the detector chamber in our experiments (by at least a factor of 50), it is assumed that $P_r$ and $S_r$ remains effectively constant.  If entropic effects are significant, the term $c\delta T$, where $c$ is the specific heat per unit mass and $T$ is temperature, should be added to the right side of Eq.~(\ref{fountainEffect}).  In all of our experiments, the entropic effects are negligible.  Eq.~(\ref{fountainEffect}) provides the basis for the differential pressure sensor acting as a mechanical spin density detector in the superfluid \hesan \aone phase.

The average deflection $\text{Z}$ of the differentia pressure sensor membrane (area $A_m$ and tension $\sigma$) is related to the differential pressure $\delta P$. If an external force $F_e$ is applied:
\begin{equation}\label{deflection}
    8\pi\sigma \text{Z} = A_m \delta P + F_e.
\end{equation}
If $F_e$ is known, $\text{Z}$ gives a direct measure of the differential pressure.  Eq.~(\ref{deflection}) assumes that, under deflection, the cross section of the circular detector membrane is  parabolic.

The superleak consists of a stack of $n$ channels each of width $w$, length $L$ and height h $\ll w,L$.  Since it is imperfect, the differential pressure produces a small concurrent normal fluid flow (with velocity $v_n$) such that:
\begin{equation}\label{normalFlow}
    \frac{\delta P}{\rho} = - GLv_n,
\end{equation}
where $G = 12\eta/\rho_n\text{h}^2$, $\rho_n$ is the normal component density, and $\eta$ is the normal component shear viscosity.

The total mass flow in the superleak is related to the membrane deflection by the conservation of mass:
\begin{equation}\label{massCons}
    \rho A_m \dot{\text{Z}} = (\rho_s v_s + \rho_n v_n) A,
\end{equation}
where $\rho_s$ is the superfluid component density, $v_s$ the superfluid component velocity and $A = nw\text{h}$ is the total cross sectional area of the superleak.

Finally, changes in $\delta S$ can be generated by a flow of the spin-polarized superfluid component flow in and out of the detector chamber.  If the superflow were the only source of change, $\delta S$ and $\delta P$ should become constant when the superflow ceases, since the balance condition expressed by Eq.~(\ref{fountainEffect}) is established under a constant applied $\delta H$.  In our experiments, $\delta P$ is always observed to decay to zero.  To incorporate a phenomenological description of this relaxation, the effects of spin density relaxation and spin diffusion are added to the net change in $\delta S$.  The net rate of change in spin density difference is written as:
\begin{equation}\label{spinRelax}
    \delta \dot{S} = \frac{\rho_s\hbar A}{2mV}v_s + \left(\frac{\chi}{\gamma}\delta H - \delta S\right)\frac{1}{\text{T}_1},
\end{equation}
where $V$ is the volume of the detector chamber and T$_1$ is the spin relaxation time.  The spin current $S_n v_n = (\chi\delta H/\gamma)v_n$ contributed by the normal fluid flow is negligibly small.  Eq.~\ref{spinRelax} ensures that the membrane dynamics are coupled to those of the spin density and the normal fluid flow.

Eq.~(\ref{fountainEffect})$\sim$(\ref{spinRelax}) are a closed set of equations governing the time ($t$) dependent response of the detector membrane to externally applied $\delta H$ and $F_e$.  The response $\text{Z}(t)$ to a step change in $\delta H$ (keeping $F_e$ = 0) is a simple exponential function with a time constant $\tau$ given by:
\begin{equation}\label{Eq_timeConstant}
    \frac{1}{\tau} = \left(\frac{1}{\tau_n} + \frac{\alpha}{\text{T}_1}\right)\left(\frac{1}{\rho_n/\rho + \alpha}\right),
\end{equation}
where
\begin{equation}\label{Eq_tau_n}
\tau_n = \frac{A_m^2\rho LG}{8\pi\sigma A}
\end{equation}
is the normal fluid flow relaxation time and
\begin{equation}\label{Eq_alpha}
\alpha = \frac{32\pi\sigma\chi m^2V}{\hbar^2\gamma^2\rho\rho_n A_m^2}
\end{equation}
is the mechanical to magnetic energy density ratio.  Since $\rho_s$ is quite small in the \aone phase, $\rho_n$ may be approximated by $\rho$.  If a step force $F_e$ is applied to the membrane instead of step magnetic field gradient, the MSE is observed.  The response is again a simple exponential with the \emph{same} time constant as given by Eq.~(\ref{Eq_timeConstant}). From the measured relaxation time $\tau$, the spin relaxation time T$_1$ is extracted using Eq.~(\ref{Eq_timeConstant}).

The ratio $\alpha$ can be evaluated from the known cell dimensions and liquid parameters \cite{Wheatley75}.  It can also be determined empirically as follows.  Let us suppose that $\delta H = 0$ and that $F_e$ has been applied for a sufficiently long time for equilibrium to be established between the detector chamber and the reservoir; hence $\delta P$ = 0 (and $\delta S$ = 0).  When $F_e$ is removed at $t$ = 0, $\delta P$ begins to develop and the superfluid accelerates out of the detector chamber.  The superflow out of the chamber is a totally spin-polarized flow.  Let $\text{Z}_0$ be the initial membrane displacement before $F_e$ is removed, and let $\text{Z}_1$ be the displacement at the time $t$ = $t_1$ when the acceleration ceases and a quasistatic equilibrium is established.  During a short time period ($\ll \tau$), if the normal flow and spin relaxation are both negligibly small, $\delta S = (\rho\hbar A_m/2mV)(\text{Z}_1 - \text{Z}_0)$ and $\delta P = -(\rho\hbar\gamma^2/2m\chi)\delta S$ at $t = t_1$.  Using Eq.~(\ref{deflection}) to relate $\delta P$ to $\text{Z}$, it can be shown
\begin{equation}\label{Eq_Z1}
    \text{Z}_1 = \frac{1}{1 + \alpha}\text{Z}_0.
\end{equation}
This relation is used to determine $\alpha$ from measurements of $\text{Z}_0$ and $\text{Z}_1$.

In the normal and A$_2$ phases, the broken relative spin-gauge symmetry on which Eq.~(\ref{fountainEffect}) is based is no longer applicable and MFE is not expected to be present \cite{Liu79} in accordance with experiments \cite{Ruel83}.  The appearance of the MFE and/or the MSE serves as clear markers for the presence of the \aone phase.

\section{experiments}
\subsection{Motivation for improving the previous apparatus}
As evident from the previous section, the relaxation time $\tau$ plays a key role to understanding the magnetically driven superflow and magnetic fountain pressure effects in the \aone phase.  In our recent study of MFE, the measured values of $\tau$ decreased towards zero with a peculiar, possibly extrinsic, kink (see Fig. 2 of Ref. \cite{Yamaguchi07}) in its temperature dependence as \tctwo was approached.  Since establishing the intrinsic temperature dependence of $\tau$ is very important and it may affect our conclusion on the existence of a minority spin condensate, it is imperative to investigate the kink in the temperature dependence of $\tau$.

\begin{figure}
\includegraphics[width=3.5in]{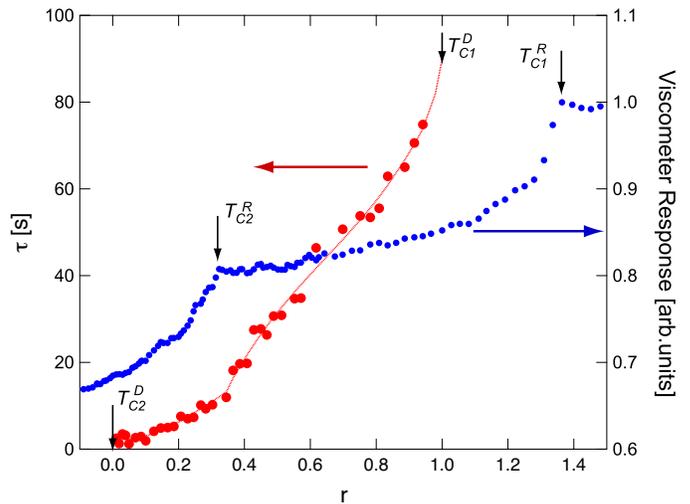}\\
\caption{(color online) Relaxation time $\tau$ vs. normalized reduced temperature $r$ (large red circles; see text) and viscometer oscillation amplitude (small blue circles).  These data were taken during a slow warming period.  The coincidence of the kink in $\tau$ and $T_{C2}^R$ at $r\thickapprox$ 0.35 is clearly evident.  P = 21 bar, $\mu_0 H$ = 8 T ($\mu_0$ is the permeability of free space).}\label{LT25fig2}
\end{figure}

To study the possibility of a thermal gradient across the superleak as the source of the observed kink in $\tau$, a new apparatus was constructed in an essentially identical manner to the previous one \cite{Yamaguchi07} using the epoxy Stycast 1266 as the construction material (see below). In addition two vibrating wire viscometer thermometers were installed; one in the detector chamber and the other in the reservoir \cite{Yamaguchi09a}.  The kink (at $r \thickapprox 0.35$) in the temperature dependence of $\tau$ was indeed reproduced in the new apparatus as shown in Fig.~\ref{LT25fig2} where $r \equiv(T - T_{C2}^D)/(T_{C1}^D - T_{C2}^D)$ is a normalized reduced temperature.  Figure~\ref{LT25fig2} also shows the simultaneously measured oscillation amplitude of the vibrating wire placed in the reservoir.  The phase transition temperatures, $T_{C2}^D$ and $T_{C1}^D$, in the detector region, were defined by the appearance and disappearance of the MFE signal, respectively, as the sample warmed.  The presence of the \aone phase (between $T_{C2}^R$ and $T_{C1}^R$) in the reservoir can be clearly identified by the kinks in the viscometer oscillation amplitude.  Note that the MFE appears at an earlier time than $T_{C2}^R$.  This shows that the liquid within the detector chamber is in fact warmer than in the reservoir.  Furthermore, the kink in the observed temperature dependence of $\tau$ coincides with $T_{C2}^R$.  Evidently, the kink occurs when the liquid in the reservoir chamber makes a transition into the \aone phase.  The temperature gradient likely occurs within the superleak channels.  The interface between the \aone phase and the \atwo phase then advances along the superleak from the detector volume side to the reservoir volume as the experiment shown in Fig.~\ref{LT25fig2} progresses.  This A$_1$-A$_2$ interface might be an interesting object for study in its own right \cite{Grabinski89,Bais90}.  However, it is now clear that the temperature gradient needs to be reduced for proper measurement of the temperature dependence of $\tau$.  The next section describes how this was accomplished.

\subsection{Improved apparatus}
\begin{figure}
\includegraphics[width=3in]{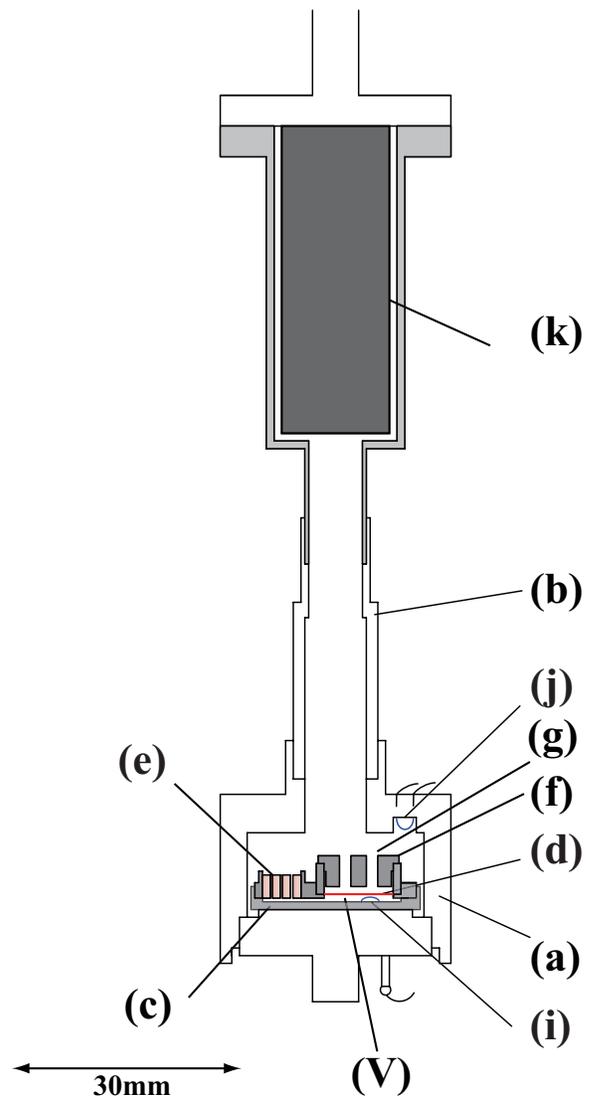}\\
\caption{Schematic of new apparatus.  The detector chamber and the \hesan container were fabricated from a machinable Macor ceramic. (a) liquid \hesan container, (b) interconnecting tower leading to the nuclear demagnetization stage, (c) detector chamber body, (d) movable membrane, (e) superleak, (f) stationary electrode holder, (g) vent holes, (i) inner vibrating wire, (j) outer vibrating wire, (k) heat exchanger, and (V) detector chamber volume. The superleak, the distance between the movable membrane and the stationary electrode holder, and the detector chamber volume are shown enlarged for clarity.}\label{apparatus_schematic}
\end{figure}

The Stycast 1266 epoxy used for fabricating the parts of apparatus was suspected as the source of long term heat release and thermal gradients \cite{Schwark85}.  With this in mind, almost all components of the new apparatus as shown in Fig.~\ref{apparatus_schematic} were reconstructed using the machinable ceramic Macor as the fabrication material.  Stycast 1266 and 2850 were used sparingly only for gluing some of the parts together.  These efforts apparently paid off, since the temperature gradient effects observed using the previous apparatus are now essentially non-existent as described below.

The liquid \hesan contained in volume (a) in Fig.~\ref{apparatus_schematic} is linked via a liquid column in an 8 mm inner diameter interconnecting tower (b) to a sintered heat exchanger in good thermal contact with a powerful copper nuclear demagnetization cooling stage.  All of the liquid \hesan associated with the experiment is exposed to an external static magnetic field applied along the vertical axis of the apparatus with uniformity better than 99 \% over the entire liquid volume.  The temperature is measured using a calibrated \hesan melting curve thermometer located in a low magnetic field region and in good thermal contact with the liquid.

The detector chamber body (c) is first assembled by gluing in the differential pressure sensor membrane (d) while leaving the superleak port open.  Before the superleak (e) is inserted into the port, the chamber body is leak-tested at 77 K to verify that there is no undesirable ancillary opening between the interior of the chamber and the outside.  The superleak is fabricated from a mold by first sandwiching 3 (= $n$) of 18 $\mu$m (= h) thick sheets of aluminum foil between thin Macor plates (affixed to the mold) of 3 mm in width and 3 mm (= $L$) in length. The aluminum is then etched away, leaving a total open cross sectional area ($A$) of 1.6$\times$10$^{-3}$ cm$^2$.  The flow impedance of the superleak was measured separately at room temperature by a gas flow test.  The differential pressure sensor membrane is a 6 $\mu$m thick circular Mylar sheet coated with aluminum film on one side.  Deflections of the membrane in response to the differential pressure between inside the detector chamber and the reservoir are detected by measuring the capacitance between the aluminum film electrode and a stationary electrode (f).  Vent holes (g) with low flow impedance bored into the stationary electrode holder equalize the pressure just above the membrane to that in the reservoir.  The active area of the Mylar membrane is 0.567 cm$^2$ (= $A_m$). The diameter of the stationary electrode is 8.5 mm.  The measured ambient capacitance ($C_0$) of the differential pressure sensor at 20 mK is 17.2$\times$10$^{-12}$ F.  The ambient average separation ($d_0$) between the membrane and the stationary electrode is thus estimated to be 29 $\mu$m.  The displacement $\text{Z}$ of the membrane is simply related to the measured change in capacitance $\delta C$ by $\text{Z} = (\delta C/C_0)d_0$.  The estimated volume ($V$) of the detector chamber is 0.13 cm$^3$.  The tension ($\sigma$) of the membrane is determined to be 2.1$\times$10$^5$ dyne/cm by measuring the changes in capacitance in response to applied voltages between the electrodes in liquid \hesan at $\sim$4 mK.

Magnetic field gradients required for observing the MFE are produced by driving currents into a set of coils designed to produce a field that varies linearly along the axis over the superleak region.  The magnetic field gradient produced by the coils was measured using a Hall probe to be 26 G/Acm. The gradient coils were wound from a superconducting wire whose critical field was about 8 T.  The critical field limited the highest static field at which the MFE could be studied.  The MSE, on the other hand, is not limited in this way and could be studied up the highest static field of 14 T.

\subsection{Electrostatic drive}
A pressure gradient for observing the MSE is produced by applying an external voltage $V_e$ between the two electrodes of the differential pressure sensor.  The applied voltage exerts a force, $F_e= C_0V_e^2/2\epsilon A_m$ on the membrane, where $\epsilon$ is the permittivity of liquid $^3$He.  The deflection that is generated can be measured by monitoring the change in capacitance.  The driven motion of the membrane acts as a mechanical spin pump which moves the spin-polarized superfluid component of the \aone phase into or out of the chamber through the superleak.  Although induced changes in spin-polarization are small in this apparatus, the changes are sufficiently large to be used for measuring relaxation processes.  As stated earlier, the MSE has an advantage over the MFE in that there is no issue with the critical current in the gradient field coils under high static magnetic fields.  In a separate MSE device specifically designed to boost spin polarization, changes in polarization greater than those observed here by four orders of magnitude have been achieved \cite{Yamaguchi09b}.

\subsection{Normal liquid \hesan flow through the superleak channels}
Prior to carrying out the MFE and MSE measurements, the characteristics of the superleak and the motion of the differential pressure sensor membrane are verified by observing the viscous flow of normal liquid \hesan through the superleak channels when subjected to a pressure difference.  Equilibrium in the membrane deflection is first established for a given applied voltage $V_e$ and liquid temperature.  Subsequently, $V_e$ is removed and $\delta C$ (and hence $\text{Z}$) is monitored as a function of time.  Apart from a small initial deviation (discussed in subsection F), the decay of $\text{Z}$ is accurately exponential (see Fig. \ref{signal}) with a time constant $\tau_n$.

\begin{figure}
\includegraphics[width=3in]{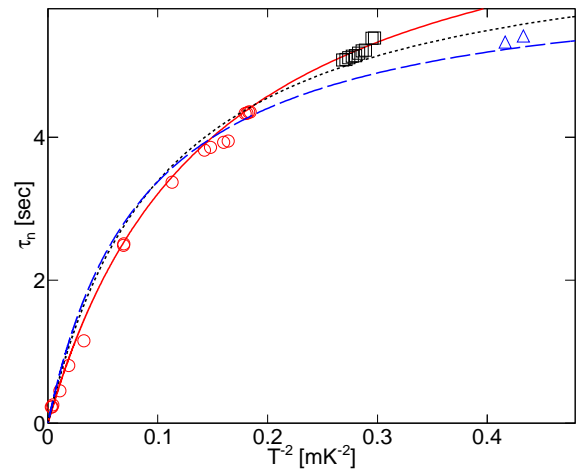}\\
\caption{(color online) Normal liquid \hesan flow relaxation time $\tau_n$ vs. $T^{-2}$.  Liquid pressures are 5 (triangles, dashed line), 10 (squares, dotted line) and 21 (circles, solid line) bar.  The curves represent fits (see text) to a simple theory that takes into account the slip effect at the boundary (see tex).}\label{TauN}
\end{figure}

Measured values of $\tau_n$ at several liquid pressures are plotted against $T^{-2}$ in Fig.~\ref{TauN}.  In the hydrodynamic regime at relatively high temperatures, $\tau_n$ is expected to be proportional to $\eta$, which varies $\propto T^{-2}$, in agreement with the data at 21 bar shown in  Fig.~\ref{TauN}.  At low temperatures, where the mean free path length approaches the superleak channel height h, the hydrodynamic flow is modified by slip effects at the boundaries.  The curves drawn in Fig.~\ref{TauN} show the expected behavior of $\tau_n$ from the ``simple'' slip effect theory \cite{Jensen80} applied to rectangular channels.  The inputs to the theory are the measured tension, the tabulated shear viscosity \cite{Wheatley75}, various cell dimensions as fabricated and a value for the channel height h adjusted to 17.7 $\mu$m.  This adjusted value is very close to the thickness of the aluminum foil used in the construction (see above) of the superleak channels.

\subsection{Magnetic fountain effect (MFE)}
\begin{figure}
\includegraphics[width=3in]{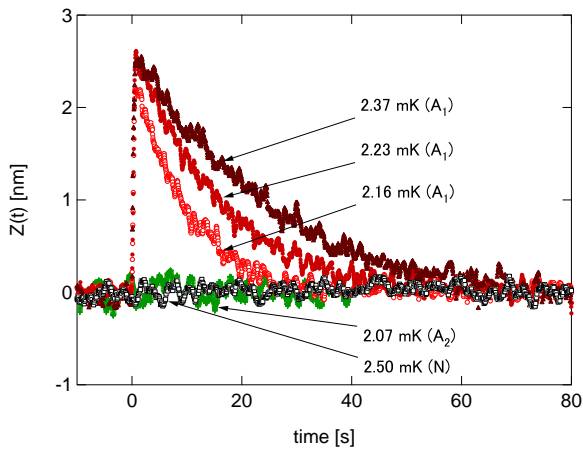}\\
\caption{(color online) Typical membrane displacement responses during MFE experiments in the A$_2$, A$_1$ and N phases (P = 21 bar, $\mu_0 H$ = 8 T).  The field gradient is linearly ramped up from -10.4 G/cm to +10.4 G/cm between $t$ = 0 and 0.2 s.}\label{MFRes}
\end{figure}

After verifying the superleak characteristics, a static magnetic field is applied for measurements in the \aone phase.  A measurement run is typically started by cooling below \tctwo into the \atwo phase.  During the subsequent slow warming of the sample caused by the residual heat leak, a series of measurements of the MFE is acquired until the liquid enters the normal phase above \tcone.  Examples of data from a typical run are displayed in Fig.~\ref{MFRes}, where the membrane displacement $\text{Z}$ derived from $\delta C$ are shown.  Here, the field gradient across the superleak is increased at $t$ = 0 from $-$10.4 G/cm to +10.4 G/cm during a time interval ($\sim$0.2 s) that is short compared to the relaxation time $\tau$.  The field gradient is kept constant for a sufficient length of time for steady-state to be reached.  The field gradient is then decreased back to $-$10.4 G/cm over the same time interval as the initial increase.  The same sequence of changes in field gradient is repeated throughout the run.  As expected, no response is observed in the \atwo (2.07 mK) and N (2.50 mK) phases where the MFE is absent.  Within the \aone phase, the influence of the MFE is clearly seen in the response of the membrane to the superfluid motion induced by the changes in the applied field gradient.  Characteristically, $\text{Z}$ reaches a peak ($\text{Z}_{max}$) just after the change in field gradient and then decays exponentially with time constant $\tau$ (see Fig.~\ref{MFTau4}).

\begin{figure}
\includegraphics[width=3.0in]{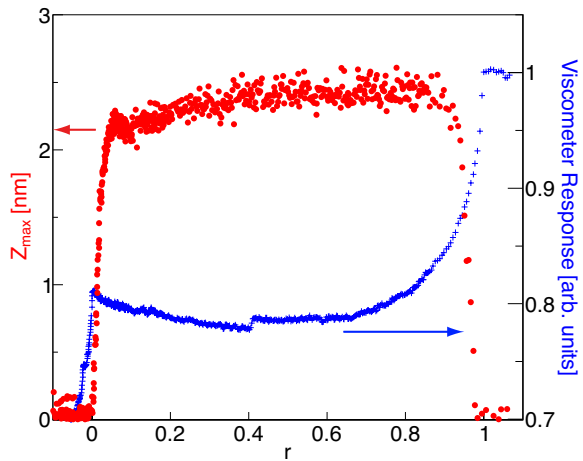}\\
\caption{(color online) Peak amplitudes of the membrane displacement and vibrating wire viscometer responses measured during the same run as shown in Fig.~\ref{MFRes}.  Coincidence of the phase boundary temperatures \tctwo and \tcone for the \aone phase as indicated by the viscometer amplitude and the MFE signal is clearly demonstrated.}\label{FigXmacorCell3}
\end{figure}

The critical measure of success in reducing temperature gradients in the new apparatus is to observe the coincidence of the appearance and disappearance of the MFE signal with the indications of \tctwo and $T_{C1}$ given to the vibrating wire viscometer in the reservoir.  Figure \ref{FigXmacorCell3} shows both the normalized viscometer response amplitude and the peak membrane displacement amplitude of $\text{Z}$ for the same data as shown in Fig.~\ref{MFRes}.  The abrupt increase and decrease in the response amplitude at \tctwo and $T_{C1}$ coincide within $\sim$5 $\mu$K to the temperatures at which the kinks in viscosity occur.  Similar coincidences are observed at all other applied static fields.  It is clear that thermal gradients in the improved apparatus are much reduced from those in the previous apparatus (cf. Fig.~\ref{LT25fig2}) and that it can be said: $T_{C1}^D = T_{C1}^R \equiv T_{C1}$ and $T_{C2}^D = T_{C2}^R \equiv T_{C2}$.  Care, however, still had to be exercised in limiting the excitation level in the capacitance bridge circuit.

It follows from Eqs. (\ref{fountainEffect})$-$(\ref{spinRelax}) that
\begin{equation}\label{}
    \delta\dot{H} = a \text{Z} + b \dot{\text{Z}},
\end{equation}
where
\begin{equation}
a = \frac{\hbar\gamma A \rho_n 8\pi\sigma}{2m\chi V G\rho A_m L}+\frac{16m\pi\sigma}{\rho\hbar\gamma A_m \text{T}_1},
\end{equation}
and
\begin{equation}
b = \frac{\hbar\gamma\rho A_m}{2m\chi V}+\frac{16m\pi\sigma}{\rho\hbar\gamma A_m}
\end{equation}
are constants.  To mimic our experiment, let $\delta H = c t$ when $0 < t < t_0$ and $\delta H = c t_0$ when $t > t_0$, where $c$ and $t_0$ are constants.  If $t_0 \ll (b/a) = \tau$, it can be shown that $\text{Z}(t = t_0) = (c/b)t_0$.  Thus the peak membrane amplitude is expected to be independent of temperature and applied field if $t_0$ is held constant.  Putting in the cell parameters to evaluate $b$ and $c$, and setting $t_0=$ 0.2 s gives $\text{Z}(t = t_0)$ = 2.7 nm.  Figure \ref{FigXmacorCell3} shows that the peak displacement is indeed comparable to this estimate.  As expected, the peak displacement is independent of temperature except near $T_{C1}$, where critical velocity effects are likely present, and near $T_{C2}$, where $\tau$ becomes comparable to $t_0$.

\subsection{Mechano-spin effect (MSE)}
\begin{figure}
\includegraphics[width=3.0in]{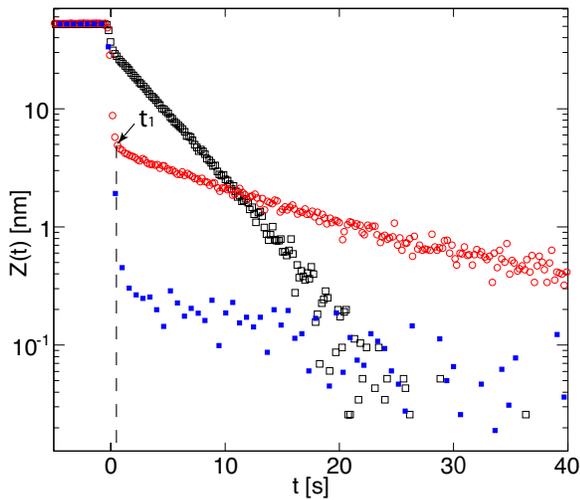}
\caption{(color online) Examples of membrane displacement response after removing an applied voltage.  A typical time sequence is as follows: $V_e$ = 30 V during -30 s $< t <$ 0 s, 0 V during 0 $\leq t< 100$ s.  Open black squares: normal liquid at $T$ = 2.7 mK, closed blue squares: \atwo phase and open red circles: \aone phase, all at P = 21 bar.  For $t \geq t_1$, the \aone phase response becomes quasistatic (see text).}
\label{signal}
\end{figure}

Let us now turn to the membrane response during MSE experiments carried out \emph{in the absence of applied magnetic field gradients}.  Initially, $V_e$ is applied for a sufficient time interval (30 s) for the membrane to come to equilibrium.  $\text{Z}(t=0)$ is set by the membrane tension, $|F_e|$, and the conditions, $\delta P$ = 0 and $\delta S$ = 0.  $V_e$ is then rapidly (within 0.2 s) reduced to zero at $t\equiv$ 0.  This sequence is repeated (usually 10 times) so that the signal can be averaged.  Typical responses to this electrostatically actuated superflow, or spin pumping, are shown in Fig.~\ref{signal} for the N, \aone and \atwo phases.  In the N phase, the response is determined by the normal fluid flow, and $\text{Z}$ decays exponentially except for an initial transient.  This small initial non-exponential decrease of the membrane shape is present at all temperatures and is likely caused by some rearrangement in the membrane shape when the pressure source changes from being electrostatic to hydrostatic.  The temperature dependence of the relaxation time $\tau_n$ in the N phase is shown in Fig.~\ref{TauN} and has already been discussed.  In the \atwo phase, $\text{Z}$ decays very rapidly to the noise floor.  The rapid decay is likely limited by some critical flow effect within the superleak.  Critical flow effects in superfluid \hesan are complicated \cite{Dahm80} and were not studied here in detail.

The time response of the membrane displacement $\text{Z}(t)$ in the \aone phase shown in Fig.~\ref{signal} is clearly distinct from those in the N and \atwo phases.  There is an initial ($t<t_1$) rapid decrease in $\text{Z}$ similar to that observed in the \atwo phase, but $\dot{Z}$ abruptly changes at a specific time which we label $t_1$ defined above.  When $t \geqq t_1$, the chemical potential is equalized across the superleak leading to a quasistatic response.  $\text{Z}(t)$ at $t > t_1$ is well-described by exponential decay from which $\tau$ (see Fig.~\ref{MFTau4}) is extracted.

\begin{figure}
\includegraphics[width=3.0in]{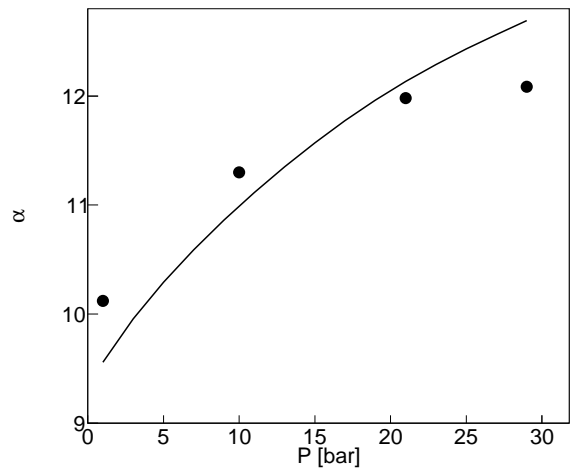}
\caption{Empirical determination of the magnetic to mechanical energy ratio $\alpha$. The line is the expected pressure dependence from Eq.~(\ref{Eq_alpha}) multiplied by 3.3.}
\label{fig-alpha}
\end{figure}

Substituting the geometric parameters of the detector, the measured membrane tension, and the liquid parameters at P = 21 bar into Eq.~(\ref{Eq_alpha}) gives the magnetic to mechanical energy ratio $\alpha$ = 3.68 dyneGauss$^2$s$^4$/g$^2$.  The empirical value of $\alpha$ determined from $\text{Z}(0)$ and $\text{Z}(t_1)$ (see Eq.~(\ref{Eq_Z1})) is larger than this by a factor of 3.3.  The difference might arise from the assumption of a parabolic membrane shape under both hydrostatically and electrostatically applied pressures.  However, the measured pressure dependence is in fair agreement with the expected pressure dependence ($\propto\chi/\rho^2$) as shown in Fig.~\ref{fig-alpha}.  At a given pressure, $\alpha$ is independent of temperature, as expected, within $\pm$5 \%.

It is interesting to compare $\text{Z}(t_1)$ with the peak displacement amplitude during the MFE experiment shown in Fig.~\ref{MFRes}.  $\text{Z}(t_1)$ is observed to be independent of temperature as expected.  The magnitude of $\text{Z}(t_1)$ is equivalent to a change in magnetic field gradient of $\sim$ 28 G/cm during similar experiments to those shown in Fig.~\ref{MFRes}.  The change in spin density $\delta S$ that would be equivalent to the pressure difference $\delta P(t = t_1)$ (cf. Fig.~\ref{signal}) using Eq.~(\ref{fountainEffect}) is only 0.02 \% of the spin density polarization ($=\chi H/\gamma$) produced by the applied field of $\mu_0 H$ = 8 T.  The change in spin density $\delta S$ induced by spin pumping here is very small owing in part to the relatively large detector volume.  The cooling expected from entropy dilution, $\sim T_{C1}A_m \text{Z}(t_1)/V$, is only several nK and is negligible.

\section{spin relaxation}
\subsection{Measurements}
A striking feature of the MFE data shown in Fig.~\ref{MFRes} is the exponential decay of the induced magnetic fountain pressure.  Equation (\ref{fountainEffect}) does not imply that this decay in $\delta P$ should occur.  A phenomenological description of this decay was introduced above in terms of spin relaxation characterized by T$_1$ (cf. Eq.~(\ref{Eq_timeConstant})).  The improvements made to the present apparatus (elimination of thermal gradients) enables to measure more accurately the temperature dependence of the relaxation time.

It is known that the \hesan spin relaxation rate in the mK temperature range is often influenced by magnetic interactions at wall boundaries \cite{Leggett70,Hu96}.  It was found in our previous MFE experiments \cite{Yamaguchi06} that coating all surfaces in contact with the sample \hesan in the apparatus with five monolayers of \heyon has no significant effect on the measured relaxation time.  Though such \heyon surface coating experiments were not carried out during the present work, we expect that they would be equally ineffective insofar as spin relaxation is concerned.

\begin{figure}
\includegraphics[width=3in]{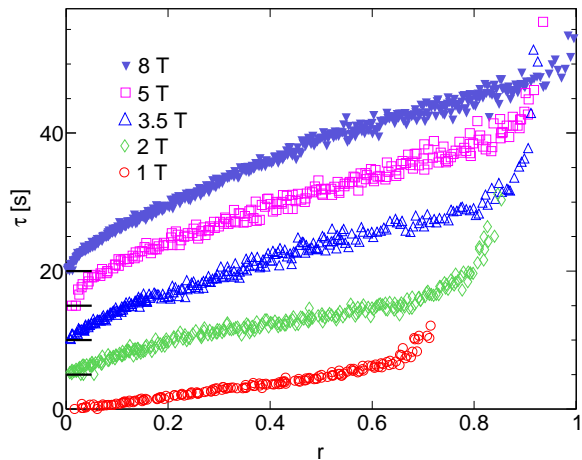}\
\caption{(color online) Values of $\tau$ extracted from MFE measurements vs. normalized reduced temperature at P = 21 bar and applied magnetic fields of 1 (circles), 2 (diamonds), 3.5 (upward triangles), 5 (squares) and 8 (downward triangles) T.  Smooth variations in $\tau$ at all temperatures and fields are evident.  For clarity, the data at 2 T and higher fields are shifted upwards by 5, 10, 15 and 20 s, respectively.} \label{MFTau4}
\end{figure}

Systematic measurements characterizing the MFE similar to those shown in Fig.~\ref{MFRes} were made as functions of applied static magnetic field and liquid pressure.  Extracted decay times $\tau$ from the measurement are shown in Fig.~\ref{MFTau4}, where the temperature is expressed as a normalized reduced temperature $r$ for convenience in comparing data acquired under different applied static fields.  Note the shifted ordinate scales for different applied fields.  The kink in the temperature dependence of $\tau$ observed previously near $r\sim$ 0.4 (see Fig.~\ref{LT25fig2}) is now absent.  It is concluded that the kink was an artifact resulting from the inadvertent presence of temperature gradients across the superleak.  Under all applied fields $\tau$ decreases monotonically towards zero as \tctwo is approached.  The previously observed tendency of $\tau$ to vanish as $T$ decreases towards $T_{C2}$ is confirmed in the present apparatus.  Thus, the observation on which the conclusion for the presence of minority spin condensate was based, still remains valid.  On the reduced normalized temperature scale, $\tau$ tends to increase more rapidly at lower fields as \tcone is approached.  The overall dependence of $\tau$ with $r$ varies weakly on the applied static field.

\begin{figure}
\includegraphics[width=3in]{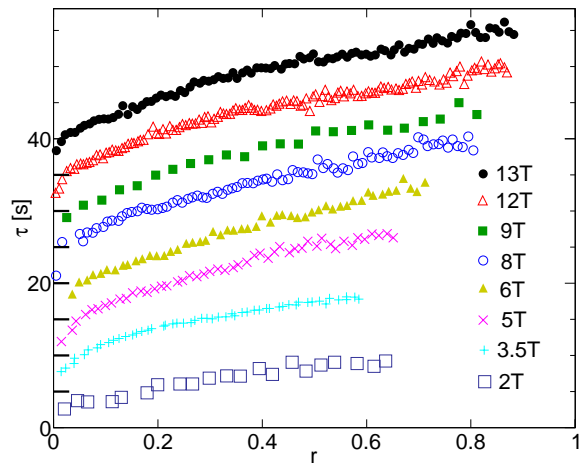}\\
\caption{(color online) Values of $\tau$ extracted from MSE measurements vs. normalized reduced temperature at P = 21 bar and applied magnetic fields of 2 (squares), 3.5 (pluses), 5 (crosses), 6 (closed triangles), 8 (circles), 9 (closed squares), 12 (open triangles) and 13 (dots) T.  For clarity, $\tau$ at 3.5 T and higher fields are shifted up by 5 s relative to the preceding dataset.}
\label{MSPTau2}
\end{figure}

Figure \ref{MSPTau2} shows $\tau$ as measured using by the pneumatically driven MSE method is shown under magnetic fields up to 13 T and a liquid pressure of 21 bar.  It can be seen that the dependence of $\tau$ on $r$ changes somewhat at low magnetic field but becomes independent of field when $\mu_0 H \geq$ 5 T.  The manner in which $\tau$ increases near \tctwo is slightly different here than in in Fig.~\ref{MFTau4}.

\begin{figure}
\includegraphics[width=3in]{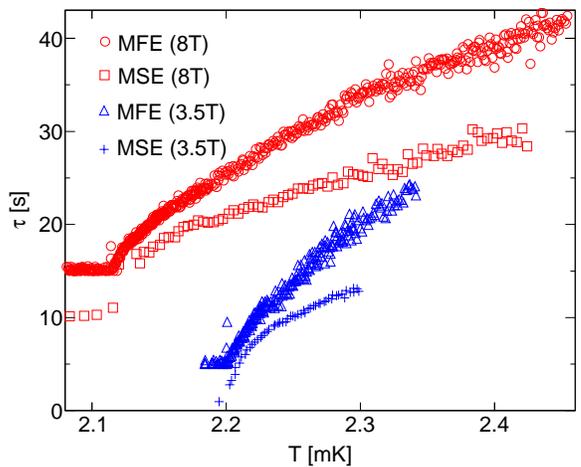}\\
\caption{(color online) Values of $\tau$ measured simultaneously using the  MFE and MSE at 8 T and 3.5 T at 21 bar and acquired during the same run.  Simultaneous measurements acquired at 2 and 5 T yield similar results.  For clarity, the data for 3.5 T (MFE), 8 T (MSE) and 8 T (MFE) are shifted up by 5 s each.}
\label{TauMFEMSP}
\end{figure}

It is expected from the simple two fluid model (see above) that the time constant extracted from the MSE (cf. Fig.~\ref{MSPTau2}) is identical to that extracted from the MFE (cf. Fig.~\ref{MFTau4}).  Many of the features of $\tau$ exhibited by data associated with these two methods are similar but not identical in detail.  To examine the apparent difference in the temperature dependence of $\tau$, ``simultaneous'' measurements of $\tau$ were acquired using both the MFE and the MSE methods during a single run as shown in Fig.~\ref{TauMFEMSP}.  The two methods were alternately applied as the temperature increased slowly.  As expected, the onset of MSE and MFE occurs at the same temperature.  The MSE method gives a steeper temperature dependence for $\tau$ near \tctwo and a more gradual one near \tconeEnd.  These differences are not likely to be caused by temperature effects since the measurements are made alternately.  With the MFE method, the applied field gradients extend over the \aone phase in the reservoir region.  Spin relaxation effects in the reservoir, which are assumed to be negligible in our simple two fluid model, \emph{might} bring about the difference.  In the case of the MSE method, the induced spin density gradients should be confined to the region in close proximity to the detector chamber itself.

\begin{figure}
\includegraphics[width=3in]{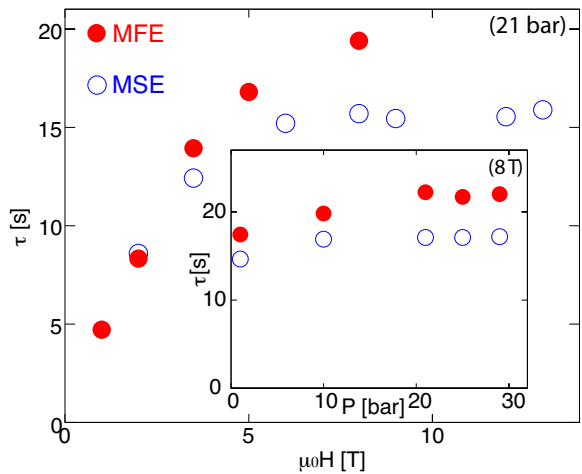}\\
\caption{(color online) Dependence of $\tau$ on applied magnetic field at $r$ = 0.5 and P = 21 bar from both the MFE and MSE measurements.  Inset shows pressure dependence of $\tau$ ($r$ =0.5, $\mu_0 H$ = 8 T).}\label{Hdep}
\end{figure}

The dependence of $\tau$ on static magnetic field at $r$ = 0.5 is shown in Fig.~\ref{Hdep}.  Up to about 4 T, both the MFE and MSE methods exhibit the same field dependence.  While the values of \tconstant measured using the MFE method continues to increase up to 8 T, those measured using the MSE method saturate in the range $\mu_0 H \gtrsim 6$ T.  Clearly, more work is needed in the high field range. Understanding of the entire field dependence of \tconstant is important to boosting the spin polarization achievable using spin pumping techniques \cite{Yamaguchi09b}.

\begin{figure}
\includegraphics[width=3in]{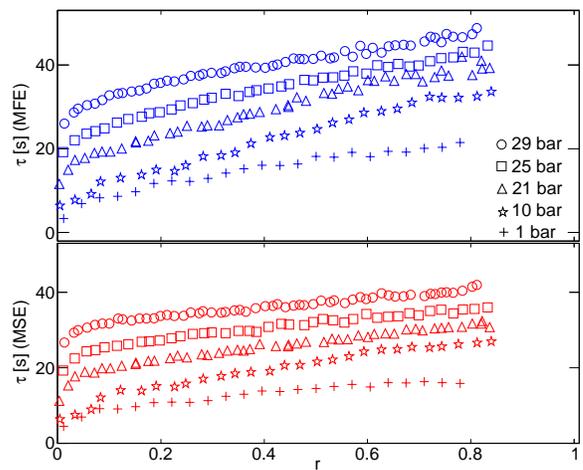}
\caption{(color online) Pressure dependence of $\tau$ as inferred from both the MFE (upper panel) and MSE (lower panel) at 8 T. For clarity, the data at 10 bar and higher are each shifted upwards by 5 s relative to the previous dataset.}\label{TauPMFEMSP}
\end{figure}

Measurements of relaxation time were also conducted using both the MFE and MSE methods at several pressures.  The results are shown in Fig. \ref{TauPMFEMSP} for an applied field of 8 T.  Both methods yield a similar weak pressure dependence of $\tau$.  The inset in Fig. ~\ref{Hdep} shows the dependence for \tconstant to pressure at $r$ = 0.5.  At pressures above about 15 bar, $\tau$ becomes independent of pressure.

\subsection{Analysis}
According to Eq.~(\ref{Eq_timeConstant}), the measured relaxation time $\tau$ depends on both the normal fluid relaxation time $\tau_n$ and the spin relaxation time T$_1$.  The quantity of more interest is T$_1$.  Since $\tau_n$ remains finite at all temperatures, the observed tendency of $\tau$ to vanish at \tctwo implies that \tone also tends to vanish there.  This surprising finding was interpreted previously \cite{Yamaguchi07} as a consequence of the presence of a minority spin condensate in the \aone phase.  The data acquired using the improved apparatus are carefully analyzed and their interpretation in terms of a model involving a minority spin condensate is reexamined.

To extract \tone from the measured values of $\tau$, the shear viscosity entering the normal relaxation time $\tau_n$ is estimated as follows.  The temperature dependence of the shear viscosity $\eta(T)$ of the \aone has been measured in high magnetic fields only at the melting pressure \cite{Roobol97} where the ratio $\eta(T)/\eta(T_{C1})$ within the \aone phase is a universal function of $T/T_{C1}$ independent of magnetic field.  For the analysis of our data, it is assumed that the same universal function gives good approximations for the temperature dependence of the hydrodynamic shear viscosity at lower pressures.  The normal fluid shear viscosity at \tcone is evaluated using a tabulation of normal fluid properties \cite{Wheatley75}.  The hydrodynamic shear viscosity is further corrected to account for slip effects \cite{Jensen80} (see discussion of $\tau_n$ above) present in the superleak channels.  The slip corrections are considerable at low temperatures and pressures where the mean free path becomes large.  However, it should be noted that the relatively large value of $\alpha$ (cf. Fig.~\ref{fig-alpha}) reduces the influence of $\tau_n$ on the value of T$_1$ from the $\tau$ data.

\begin{figure}
\includegraphics[width=3in]{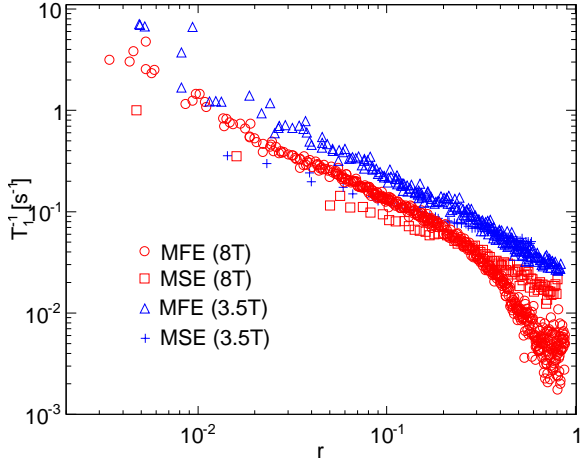}
\caption{(color online) Extracted relaxation rate T$_1^{-1}$ vs reduced temperature $r$.  The simultaneous measurements of \tconstant at P = 21 bar shown in Fig.~\ref{TauMFEMSP} are used in conjunction with Eq.~(\ref{Eq_timeConstant}) to extract T$_1^{-1}$.  Symbols are the same as those used in Fig. \ref{TauMFEMSP}.}\label{T1MFEMSP}
\end{figure}

Fig. \ref{T1MFEMSP} summarizes the spin relaxation rate T$_1^{-1}$ extracted in the manner described above from the simultaneous measurements of $\tau$ by MFE and MSE.  As it was already implicitly evident from $\tau$ (shown in Fig.~\ref{TauMFEMSP}), the dependence of \tone on $r$ derived from the MFE and MSE methods qualitatively track one another except for two slight differences: near \tctwo values of \tone extracted from the MSE data are longer than those from the MFE, and near \tcone those from the MFE data show stronger temperature dependence than those from the MSE data.  It is clear from Fig.~\ref{T1MFEMSP} that T$_1^{-1}$ continues to increase as \tctwo is approached.  The maximum relaxation rate that can be measured is limited to about 10 s$^{-1}$ by the time constant of the lock-in amplifier used for capacitance detection.  No transport properties such as viscosity or spin diffusion in the \aone phase are known to vanish or diverge at \tctwo with the possible exception of a preliminary report by Awobode and Leggett \cite{Awobode09}.  It is concluded that the large increase in T$_1^{-1}$ near \tctwo originates in an intrinsic spin relaxation process occurring in the \aone phase.

\begin{figure}
\includegraphics[width=3in]{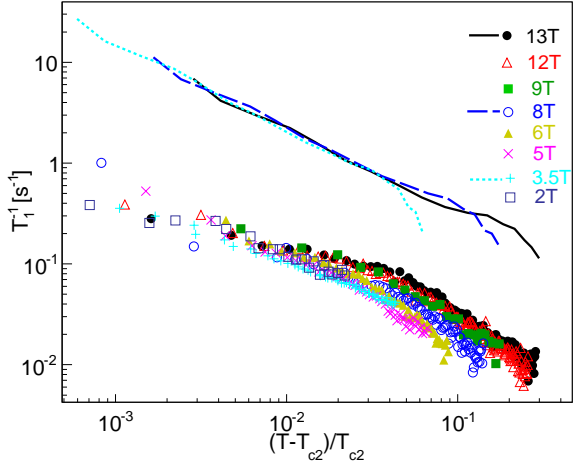}\\
\caption{(color online) Values of T$_1^{-1}$ vs. reduced temperature $(T - T_{C2})/T_{C2}$ for the data shown in Fig. \ref{MSPTau2} and the same symbols used for each applied field.  Note the change in the abscissa from the normalized reduced temperature $r$ in Fig.~\ref{T1MFEMSP}. Theoretical relaxation rates (see text) are shown by lines at 3.5 (dotted), 8 (dashed) and 13 (solid) T.}
\label{invT1}
\end{figure}

The values of T$_1^{-1}$ extracted from $\tau$ acquired via the MSE method and shown in Fig.~\ref{MSPTau2} are plotted as a function of $(T - T_{C2})/T_{C2}$ in Fig.~\ref{invT1}.  The data from the MSE method are chosen because of the greater range of magnetic field than can be applied.  Although there is some scatter, the experimentally determined relaxation rate shown in Fig.~\ref{invT1} can be simply represented by T$_1^{-1} \propto [(T - T_{C2})/T_{C2}]^{-\beta}$ with $\beta\sim$ 0.6 in the restricted temperature range close to \tctwo where $(T - T_{C2})/T_{C2} \lesssim$ 0.02.

The argument for the minority spin condensate in the \aone phase as the origin of the T$_1^{-1}$ increase near \tctwo is briefly as follows.  Monien and Tewordt (MT) \cite{Monien85} showed that the small but finite minority spin condensate emerged when the dipolar interaction energy was included in the total free energy.  The presence of both majority and minority pair condensates implied that a longitudinal magnetic resonance (with frequency $\Omega_{\parallel}$), which otherwise would be absent without the minority condensate, could occur in the \aone phase \cite{Monien85}.  The presence of a minority condensate then allows the Leggett-Takagi (LT) mechanism \cite{Leggett77} to contribute in the spin relaxation process, and consequently the spin relaxation rate can dramatically increase.

According to the LT mechanism, the spin density relaxation rate ($\Gamma_{\parallel}$) of a longitudinal magnetic resonance in the A phase is given by $\Gamma_{\parallel} = (1 - \lambda)\tau_{qp}\Omega_{\parallel}^2/2\lambda(1 + \zeta/4)$.  Here, $\lambda = 1 - Y_2(T)$, where $Y_2$ is the ``second order'' Yosida function \cite{Leggett77}, $\tau_{qp}$ the quasiparticle relaxation time \cite{Wheatley75} (to be assumed equal to that at \tconeEnd), $(1 + \zeta /4)^{-1}$ is the ratio of liquid magnetic susceptibility \cite{Wheatley75} to the ideal Fermi gas susceptibility and $\zeta$ is a Landau parameter.  In the spirit of the quasi-static treatment of our experiment, we have hypothesized \cite{LT_hypothesis} that the measured T$_1^{-1}$ be identified with $\Gamma_{\parallel}$.

MT computed the temperature dependence of $\Omega_{\parallel}(r_1)$ with $r_1 \equiv 1 - r$ and found it to be independent of magnetic field up to 2 T (cf. Fig. 5 of Ref.~\cite{Monien85}).  By computing $\lambda(T)$ for each applied field and assuming $\Omega_{\parallel}(r_1)$ is independent of field up to 13 T, the theoretical relaxation rate $\Gamma_{\parallel}$ is evaluated \emph{without} adjusting any parameters and shown by the lines drawn on Fig.~\ref{invT1}.  The temperature dependence of the theoretical $\Gamma_{\parallel}$ agrees generally with the experimentally extracted values of $\text{T}_1^{-1}$.  Insensitivity to applied field in the theory appears to be consistent with the experiment close to \tctwo but not at higher temperatures.  However, the overall magnitudes do not agree.  If the minority energy gap $\Delta_{\downarrow\downarrow}$ were reduced by a factor 16, the theoretical prediction for $\Gamma_{\parallel}$ can be brought into agreement with the experimental data for T$_1^{-1}$.  It is possible that the $\Omega_{\parallel}$ calculated by MT for the bulk \aone phase may be different than that in our finite cell geometry contributing to the discrepancy.

\begin{figure}
\includegraphics[width=3in]{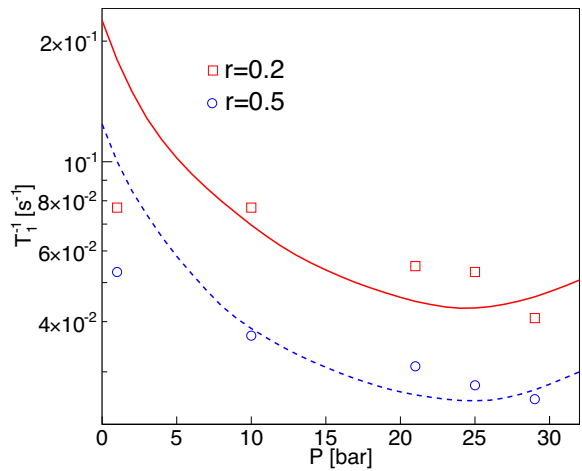}\\
\caption{(color online) Measured pressure dependence of T$_1^{-1}$ at 8 T and normalized reduced temperatures of 0.2 (red squares) and $r$ = 0.5 (blue circles). The scaled theoretical pressure dependence (see text) is shown by solid and dotted curves for $r$ = 0.2 and 0.5, respectively.}
\label{T1vsPgraph}
\end{figure}

The pressure dependence of T$_1^{-1}$ extracted from $\tau$ at $r$ = 0.5 and 0.8 under an applied field of 8 T is shown in Fig.~\ref{T1vsPgraph}.  The lines indicate the theoretical pressure dependence based on the presence of a minority spin condensate as follows.  According to MT, $\Omega_{\parallel}^2$ in the \aone phase is estimated as $\sim \gamma^2(g_D/\chi)\Delta_{\uparrow\uparrow}\Delta_{\downarrow\downarrow}$, where $g_D$ is the dipolar energy \cite{Monien85} and the minority spin condensate energy gap $\Delta_{\downarrow\downarrow} \approx (g_D/\eta^{\prime}H)\Delta_{\uparrow\uparrow}$.  The term $\eta^{\prime}H$ gives a measure of the transition temperature \tcone in magnetic field $H$ \cite{Monien85}.  It can then be shown $\Omega_{\parallel}^2 \sim (\gamma^2 g_D^2/\chi\beta_{24})(1 - r/r_c)$, where $\beta_{24}=\beta_{24}^{\prime}(21\zeta (3)/40\pi^2)(N(0)/k_B^2 T_c^2)$, $\beta_{24}^{\prime}$ is a strong coupling parameter \cite{Akimoto97,Tang91}, $N(0)$ is the density of states, $T_c$ is the transition temperature in zero field, and $r_c \equiv (T_{C1} - T_C)/T_C$.  The pressure dependence of $g_D/\chi$ can be estimated from the measured temperature dependence of the longitudinal resonance frequency in the A phase as the temperature approaches $T_c$ \cite{Schiffer92}.  The theoretical pressure dependence of $\Gamma_{\parallel}$ divided by 16 is shown by the curves drawn on Fig. \ref{T1vsPgraph}.  The observed dependence of T$_1^{-1}$ on pressure is thus roughly consistent with the theoretical expectation.

\section{Summary}
We demonstrated that temperature gradients had been present in our previous magnetic fountain effect (MFE) experiments \cite{Yamaguchi07} and that these gradients had influenced data in the ferromagnetic superfluid \hesan \aone phase.  The work presented here was motivated by the goal of eliminating these temperature gradients.  The temperature gradients were successfully eliminated by replacing almost all of the epoxy that was in contact with superfluid \hesan in the previous apparatus by machinable ceramic (Macor).  The important observations made previously \cite{Yamaguchi07} in which the spin relaxation time tended to vanish as the the A$_1$-A$_2$ phase transition temperature is approached was observed to persist when the temperature gradients were eliminated.  The detection scheme was modified to permit observations of the mechano-spin effect (MSE) where mechanical spin pumping of the spin-polarized superfluid component of the \aone phase was generated by electrostatically actuated membrane motion.  The new measurements characterizing the MSE demonstrated that the same magnetic relaxation processes could be observed without imposing magnetic field gradients as required by the MFE experiments.  The spin relaxation rate (T$_1^{-1}$) was extracted as functions of temperature, pressure and magnetic field.  The temperature dependence of the extracted rate T$_1^{-1}$ agrees well with that deduced by a formulation combining Leggett-Takagi spin dynamics with the existence of minority spin condensate as predicted by Monien and Tewordt.  Our observations call for more theoretical studies of the minority spin condensate in the \aone phase and of the exact relationship between minority spin condensate and the LT mechanism.   Experimental improvements in temperature regulation and faster response in the capacitance detection system is desirable in the future for probing the possible divergence of T$_1^{-1}$ near \tctwo.

\section{acknowledgment}
We thank A. Awobode, H. Ebisawa, W. Halperin, A. Leggett, K. Nagai and T. Takagi for discussions.  This research was supported by JSPS Grant-in-Aid Scientific Research funds(19340091 and 22684019) and by the US NSF (DMR-0704120) and INT-NSF (INT-0234032).

\bibliography{helium3}

\begin{thebibliography}{31}
\bibitem[{LT_()}]{LT_hypothesis}%
  \BibitemOpen
  \href@noop {} {}\bibinfo {note} {The longitudinal relaxation rate
  $\Gamma_{\parallel}^{-1}$ was derived [27] for small changes in magnetic field,
  whose associated energy is small compared to the dipolar energy of order
  $\chi\Omega_{\parallel}^2/\gamma^2$. The changes in energy induced by the
  applied field gradient or by the pumped spin in our experiment is
  considerably greater than the dipolar energy. Nevertheless, the relaxation
  rate the $\Gamma_{\parallel}^{-1}$ appears to be nearly applicable to our
  experiment.  The justification for this apllicability is not yet clear to us.}%
  \bibAnnoteFile{Stop}{LT_hypothesis}
\end{thebibliography}
\end{document}